# High-buckled $\sqrt{3} \times \sqrt{3}$ stanene with topologically nontrivial energy gap

Ye-Heng Song[†], Zhi-Wen Wang[†], Zhen-Yu Jia[†], Xin-Yang Zhu, Zhi-Qiang Shi, Li Zhu, Qian-Qian Yuan, Da-Jun Shu[*], Shao-Chun Li[*]

*National Laboratory of Solid State Microstructures, School of Physics, Nanjing University, Nanjing 210093, China.*

*Collaborative Innovation Center of Advanced Microstructures, Nanjing University, Nanjing 210093, China.*

[†] These authors contributed equally to this work.
Email: djshu@nju.edu.cn (D.J.S.); scli@nju.edu.cn (S.C.L.)

**Abstract**：

Stanene has been predicted to be a two-dimensional topological insulator (2DTI). Its low-buckled atomic geometry and the enhanced spin-orbit coupling are expected to cause a prominent quantum spin hall (QSH) effect. However, most of the experimentally grown stanene to date displays a metallic state without a real gap, possibly due to the chemical coupling with the substrate and the stress applied by the substrate. Here, we demonstrate an efficient way of tuning the atomic buckling in stanene to open a topologically nontrivial energy gap. Via tuning the growth kinetics, we obtain not only the low-buckled 1×1 stanene but also an unexpected high-buckled √3×√3 stanene on the Bi(111) substrate. Scanning tunneling microscopy (STM) study combined with density functional theory (DFT) calculation confirms that the √3×√3 stanene is a distorted 1×1 structure with a high-buckled Sn in every three 1×1 unit cells. The high-buckled √3×√3 stanene favors a large band inversion at the Γ point, and the spin orbital coupling open a topologically nontrivial energy gap. The existence of edge states as verified in both STM measurement and DFT calculation further confirms the topology of the √3×√3 stanene. This study provides an alternate way to tune the topology of monolayer 2DTI materials.

Two-dimensional topological insulators (2DTIs) host the quantum spin Hall (QSH) effect and feature the insulated bulk gap and gapless metallic edge states protected by time-reversal symmetry[1-4]. Due to the spin-momentum locking, back scattering is strictly prohibited in the topological edge state[3]. Such feature enables the dissipationless edge conduction and thus is promising in low consumption electronic devices. The 2DTI was first predicted in graphene[5], but the spin-orbit coupling (SOC) is so weak that the opened gap is expected to be in the magnitude of μV [5]. The QSH insulator was subsequently predicted [6] and experimentally realized [7] in the HgTe/CdTe quantum wells. Again, its small band gap (a few mV) makes the working temperature still below ~10K. Searching for large gap QSH insulator is vitally necessary for realization of high temperature QSH.

The 2D monolayered Sn in a buckled honeycomb lattice, namely stanene, has been predicted to be the candidate of 2DTIs [8,9]. Stanene takes a similar honeycomb structure as graphene, but with low-buckled atomic geometry[8,10,11]. Theoretical calculations indicated that the combination of the large atomic number and the low-buckled structure greatly enhances the SOC in stanene that results in a large band gap 2DTI [8,9,11]. Other fascinating properties [9,11-20], such as topological superconductivity [18], high-temperature QAH effect [8,16] and enhanced thermoelectricity[19], were also predicted in the stanene associated systems. Recently, the monolayer stanene has been grown on $Bi_2Te_3$(111) substrate [21]. However, the epitaxial stanene usually exhibits metallic band structure due to its coupling to the substrate and the induced compressive lattice[21,22]. To the best of our knowledge, the QSH stanene with a nontrivial band gap has not yet been realized, and the experimental verification of the topological edge states remains absent.

In this study, we successfully grew the high-quality stanene monolayer on the Bi(111) substrate. The lattice mismatch between Bi and Sn makes the stanene monolayers slightly compressed. By tuning the growth kinetics, we obtained an unexpected $\sqrt{3} \times \sqrt{3}$ (R3) structure of stanene as well as the normal 1×1 stanene. The R3 stanene is like a distorted 1×1 stanene consisting of a high-buckled Sn atom per three 1×1 unit cells. The R3 stanene is more stable than the 1×1 stanene at low temperature. The high buckling of Sn atoms in R3 favors a large band inversion at the Γ point, while the spin orbital coupling open a topologically nontrivial energy gap. The scanning tunneling spectroscopy (STS) measurements identify the edge states at the step edge of stanene, and the DFT calculations confirm their topological origin.

To grow high quality and large area Sn monolayer, the lattice match between the Sn and substrate is important[21,23,24]. The in-plane lattice constant of the Bi(111) is ~0.454 nm [25], very close to that of the predicted freestanding Sn monolayer, ~0.468 nm [8]. In addition, the coupling to Bi(111) is relatively weak because of its van der Waals layered structure. Therefore we adopt the epitaxial Bi(111) films of ~ 10 MLs thick on Si(111) as the substrate. Figure 1a shows the surface morphology of Bi(111) film and Fig. 1b the surface after directly depositing ~0.6 MLs Sn on the Bi(111) at room temperature (RT). The surface is characteristic of Sn islands with the uniform height in addition to some small clusters. The apparent height of the Sn islands is ~0.4 nm, as shown in Fig.1d, consistent with the predicted value for the freestanding Sn monolayer [8]. The atomic resolution topographic image taken on the Sn monolayer, the inset to Fig. 1b, clearly reveals the 1×1 honeycomb structure. The in-plane lattice constant of the grown 1×1 stanene is ~0.45nm and close to that of the Bi(111) substrate, indicating ~3% compressive strain in the stanene.

When Sn is deposited at low temperature followed by annealing to RT, another unexpected structure of Sn monolayer starts to appear as a result of tuning the epitaxial kinetics. The as-deposited Bi(111) surface is covered by uniformly distributed Sn clusters, as shown in the supplementary Fig S1. After annealing, the Sn clusters grow into large and high quality Sn film. Figure 1c shows the surface image of ~0.6 MLs Sn deposited on Bi(111) at an optimized low temperature followed by annealing to RT. The Sn films mainly consist of the new structure, together with a minority of the 1×1 stanene islands, as marked in the Fig. 1c. The apparent step height of the new Sn structure is ~0.35 nm and slightly smaller than that of the 1×1 stanene, as shown in the line-scan profiles in Fig. 1d. Highly resolved STM image taken on the new structure of Sn monolayer, Figs. 1e, reveals a different hexagonal periodicity with an in-plane lattice constant of ~0.79 nm, approximately $\sqrt{3}$ times of the 1×1 lattice constant. More atomic resolution STM data can be found in the supplementary Fig. S2. Furthermore, the fast Fourier transform of Fig. 1e indicates that the unit cell is rotated by 30° relative to that of 1×1 stanene, as shown in Fig. 1f. All these results suggest that the unexpected stanene monolayer is in a $\sqrt{3} \times \sqrt{3}$ structure. It is worthwhile emphasizing that the $\sqrt{3} \times \sqrt{3}$ stanene only forms at low temperature. Direct deposition of Sn on Bi(111) at RT only leads to the 1×1 stanene structure, as seen in Fig.1(b). We rule out the possibility that Sn and Bi form alloys on the surface, because the sharp interface between the Bi step edge and the epitaxial Sn monolayer is very sharp according to the dI/dV map, see supplementary Fig. S3.

DFT calculations are performed to help understand the atomic structures of the grown stanene. The in-plane lattice constant of the freestanding stanene is 4.67 Å, in agreement with the values reported previously[8]. For stanene on Bi(111) surface, two different periodicities can be obtained. Direct relaxation from the uniformly compressed stanene leads to the 1×1 structure. In combination with the experimental measurements, we affirm that the Sn atoms in the bottom and the top layer locate at the hcp and fcc hollow sites, respectively, denoted as (AB)AC stacking in Fig. 2a. See supplementary Figs. S4 and S5 for comparison of the other stacking in the supplementary materials. In contrast, a new $\sqrt{3} \times \sqrt{3}$ structure (R3) is reached upon relaxation as shown in Fig. 2b, if the Sn atoms are displaced to a certain extent away from the ideal positions. The DFT simulated images based on the R3 model, Figs. 2c and d, are greatly consistent with the experimental STM results, Figs. 2e and f. Furthermore, it is clear that the center and the three vertexes of the triangular patterns observed correspond to the Sn atoms in the topmost layer and in the bottom layer, respectively. Therefore, based on the contrast of the STM images of 1×1 and R3 in Fig. 1e and supplementary Figs. S6-S8, the R3 structure is affirmed as a distorted 1×1 in stacking of (AB)CA with a Sn atom buckled heavily upwards in every three (1×1) unit cell. The R3 structure is more stable than the 1×1 in stacking of (AB)AC by ~0.066 eV/atom. The magnitude of the vertical buckling in the R3, ~1.8 Å, is much larger than the value in the 1×1 structure, ~1.0 Å.

The projected band structures of the 1×1 and $\sqrt{3}\times\sqrt{3}$ stanene placed on Bi(111) are plotted in Fig. 3a,b. In the 1×1 stanene on Bi(111), the SOC-induced gap at $K$ point is pinned down which results in a metallic band structure, consistent with the previous observations in the 1×1 stanene grown on $Bi_2Te_3$ [21]. Instead, the projected band structure of the R3 stanene indicates a gap with the size of ~0.02 eV locating at ~0.16 eV above the Fermi level.

The differential conductance dI/dV spectra taken on the 1×1 and R3 stanene are shown in Fig. 3c,d. The dI/dV curve measured in the R3 stanene exhibits clearly a gap shape near the Fermi level. In contrast, the dI/dV curve for the 1×1 stanene shows continuous states without any identifiable gap shape. For comparison, the calculated local densities of states (LDOS) for both of the structures are also plotted. The calculated LDOS agree very well with the measured STS spectra after shifting both LDOS towards lower energy by ~0.18 eV. The agreement between the STS measurement and the calculated LDOS further confirm the calculated 1×1 and R3 structures. More importantly, it indicates that the gap-shaped feature observed in the STS for R3 stanene (near the Fermi level) corresponds to the projected gap region (at ~+0.16 eV) in the calculated band structure, Fig. 3b. This supports the existence of a band gap in the epitaxial R3 stanene. It may be the substrate contribution that induces the finite intensity of dI/dV spectra within the gap region.

In order to provide an in-depth insight of the electronic structure, we exfoliate the R3 stanene from the Bi(111) substrate and calculate its band structure while keeping the atomic positions unchanged. As shown in Fig. 4a, an obvious band inversion occurs at around the Γ point of the exfoliated R3 stanene, between the $p_z$ band in the conduction band and the $p_x + p_y$ ($p_{//}$) band in the valence band. Meanwhile a dramatic gap of ~0.1 eV appears at ~0.3 eV above the Fermi level. In contrast, the exfoliated 1×1 structure remains gapless, as shown in supplementary Fig. S5.

For comparison, the band structure of the exfoliated R3 stanene is also calculated without the inclusion of SOC, as shown in Fig. 4b. We can see that SOC enlarges the band inversion between $p_z$ and $p_{//}$ band at Γ point, and opens a topologically nontrivial gap in the R3 stanene. The high-buckled atomic geometry in the R3 stanene plays key role in opening the energy gap. Firstly, the high buckling induces the band inversion between $p_z$ and $p_{//}$ band. Secondly, it enhances the effective SOC and thus results into a large energy gap [11,20].

Next, we explore the edge states at the step edge of R3 stanene. The dI/dV spectra taken along a line crossing the step edge unveil the electronic evolution in the R3 stanene, as shown in Fig. 5a. The corresponding topographic image of the stanene/Bi(111) step is depicted in the inset to Fig. 5a. As approaching to the step edge from the stanene terrace, the main features are kept unchanged as identified in the dI/dV spectra, such as the V shape at the Fermi level. However, new peaks at +180 mV and +400 mV emerge at the step edge, indicating the existence of edge states. We further took spatially resolved dI/dV maps on the R3 stanene monolayer. As shown in Figs. 5b-d, the edge state robustly distributes along all the periphery of the R3 stanene terrace, with a penetration depth of ~2 nm. Since the momentum-space width of the edge-state dispersion scales with the inverse of the penetration depth[26], the obtained penetration depth of ~2 nm for the R3 stanene is close to the size of one unit cell, indicating that the edge states extend over the whole Brillouin zone. In contrast, the step edge of 1×1 stanene exhibits no prominent edge states, as shown in the supplementary Fig. S9.

We build a nanoribbon in width of ~13.4 nm with the exfoliated R3 stanene to calculate the edge states. As marked by the red lines in the band structure in Fig. 5e, the gapless edge states appear within the gap above the Fermi level. For the R3 stanene on the Bi(111), the electronic structure is more or less affected by the substrate, therefore the position of the calculated edge states is not directly comparable to the experimental results. Even so, the theoretical results verify the existence of the topological nontrivial edge states in R3 stanene.

In summary, by tuning growth kinetics, we obtained an unexpected R3 stanene monolayer on the Bi(111) substrate. STM measurement and DFT calculation confirmed that in the R3 stanene, Sn atoms are high-buckled, comparing to the 1×1 stanene formed on the same substrate. Such a high-buckled geometry results in band inversion and a large effective SOC, and thus opens a topologically nontrivial gap, in contrast to the gapless 1×1 stanene. As the R3 stanene on Bi(111) substrate is stable at low temperature, it is expected that such a new structure of stanene may exist in the real stanene-based device. Once the Fermi level is tuned to be within the gap, for example by electrical gating or doping, the 2D TI can be realized with the unique R3 stanene.

**Acknowledgements:**
We thank the fruitful discussion with Yong Xu and Dr. Xiangang Wan. This work was supported by the Ministry of Science and Technology of China (Grants No. 2014CB921103, No. 2013CB922103), the National Natural Science Foundation of China (Grants No. 11374140, 11174123), and the Basic Research Plan of Jiangsu Province (Grant No. BK20161390).
Y.H.S, Z.W.W, and Z.Y.J. contributed equally to this work.

**Experiments and methods:**

The sample was prepared in a standard MBE chamber connected to an Unisoku low-temperature (LT) STM system (base pressure $1\times10^{-10}$ mbar). The substrate of Bi(111) thin films were prepared by depositing high purity Bi (99.999%, Alfa Aesar) of ~10 monolayers (MLs) on the Si(111) 7×7 substrate. The Si(111)-7×7 was pre-cleaned by flash annealing to ~1300K for a few cycles. The Si(111)-7×7 substrate was kept at room temperature (RT) during the Bi evaporation. The Bi/Si(111) was then annealed to ~400 K for ~24 hours. As a result, the clean Bi (111) film with relatively large surface terraces was produced. The flux rate of Bi is ~ 0.2 MLs /min. During the RT growth of stanene, the Bi(111)/Si(111) substrate was kept at RT for RT-growth of stanene, and at low temperature (LT) for LT-growth of stanene, The Bi(111)/Si(111) substrate was cooled down to ~80 K with liquid nitrogen. After the LT deposition, the Sn/Bi(111)/Si(111) sample was then annealed to ~RT to obtain the continuous Sn film. The as-deposited Sn/Bi(111)/Si(111) surface was also quickly checked by STM right after the LT deposition and prior to RT annealing.

STM characterizations were carried out at ~78 K and ~4 K. STM images were collected with a constant current mode. All the dI/dV measurements were acquired using a standard lock-in technique with the ac modulation of 5~10 mV at a frequency of 963Hz. A chemically etched STM tip was used for scan.

The density functional theory calculations [27] were performed with the Vienna Ab Initial Simulation Package (VASP) in the PBE generalized gradient approximation [28,29] with the semi-empirical vdW correction based on Grimmes scheme (PBE-D2) [30-32]. An energy cutoff of 350 eV was used for expanding the Kohn-Sham wave functions. With these settings, the lattice parameters of bulk Bi is a = 4.531 Å and c/a = 2.608, which agrees well with the experimental values [33]. The Bi(111) surface was modeled by using a ($n\times n$) supercell consisting of a six-bilayer slab plus a vacuum with thickness of 20 Å, where $n$ is 3 in the structure relaxation. The Brillouin zone (BZ) of the (3×3) supercell was sampled by a 3×3 $\Gamma$-centered mesh in the structure relaxation. The positions of atoms in the bottom two bilayers were fixed to mimic the bulk, and the other atoms were relaxed until the forces were converged to 0.01 eV/ Å. In the local density of state (LDOS) and band calculation of the 1×1 or R3 structure, the (1×1) or ($\sqrt{3}\times\sqrt{3}$) supercell was used, and the corresponding surface BZ was sampled by an 18×18 and a 10×10 k-mesh, respectively. The STM images were simulated using the Tersoff and Hamann approximation [34]. The nanoribbon of the stanene was modeled by inserting an in-plane vacuum gap of 13 nm between slabs.

**FIRGURE CAPTION**

**Figure 1: STM topography of 1×1 and R3 stanene on Bi(111).** (a) STM image of the Bi(111) film grown on Si(111)-7×7. (200 × 200 nm$^2$, U = +1.0 V, I$_t$= 100 pA). (b) Stanene film of ~0.6 MLs growth on Bi(111) at RT (80 × 80nm$^2$, U = +1.0 V, I$_t$ = 100 pA). (c) Stanene film grown on Bi(111) at ~140K after annealing process (80 × 80 nm$^2$, U = +1.0V, I$_t$ = 100 pA). (d) Line-scan profiles measured along the lines marked in (b) and (c), showing the step heights for R3 and 1×1 stanene. (e) Atomically resolved STM image of stanene surface showing both of the 1×1 and R3 regions (12×12 nm$^2$, U= +10 mV, I$_t$ = 1.0 nA). (f) The FFT image of the atomically resolved STM image in (e), the blue and red arrows mark the vectors for the 1×1 and $\sqrt{3}\times\sqrt{3}$ stanene.

**Figure 2: DFT determined structure models of 1×1 and R3 stanene on Bi(111).** (a),(b) DFT optimized atomic structures of the observed 1×1 and R3 stanene on Bi(111), respectively. The blue dotted parallelograms in (a) and (d) mark the supercell of (1×1) and ($\sqrt{3} \times \sqrt{3}$) stanene. (c),(d) DFT simulated images of R3 stanene at various bias voltages (c: U = +0.5 V; d: U = +60 mV). (e),(f) Atomically resolved STM images of R3 stanene (e: U = +0.5 V, I$_t$ = 100 pA; f: U = +10 mV, I$_t$ = 100 pA).

**Figure 3: Electronic structure of 1×1 and R3 stanene.** (a) and (b) Projected band structure of the 1×1 stanene and R3 stanene placed on the Bi(111) substrate. (c) and (d) Typical differential conductance (dI/dV spectra) taken on the 1×1 and R3 stanene terrace compared with the corresponding projected LDOS of the stanene.

**Figure 4: Band structures of the exfoliated R3 stanene.** The spin-orbital coupling is included in (a) and not included in (b).

**Figure 5: Edge states in R3 stanene.** (a) dI/dV spectra taken along a line across the R3 stanene step edge and Bi(111) substrate. The black curves are measured on the stanene terrace and blue on the Bi(111) substrate. The red triangles mark the features for the edge states. (b-d) STM topographic image (38×25 nm$^2$, U = +400 mV; I$_t$ = 200pA) of the R3 stanene island and the corresponding dI/dV maps taken at +180 mV (c) and +400 mV (d) respectively. The black curves outline the step edge of the R3 stanene island as determined in (b). The R3 stanene is in connection with the 1×1 stanene in the right-upper corner. (d) Band structure of the nanoribbon in width of ~13.4 nm of the exfoliated R3 stanene.

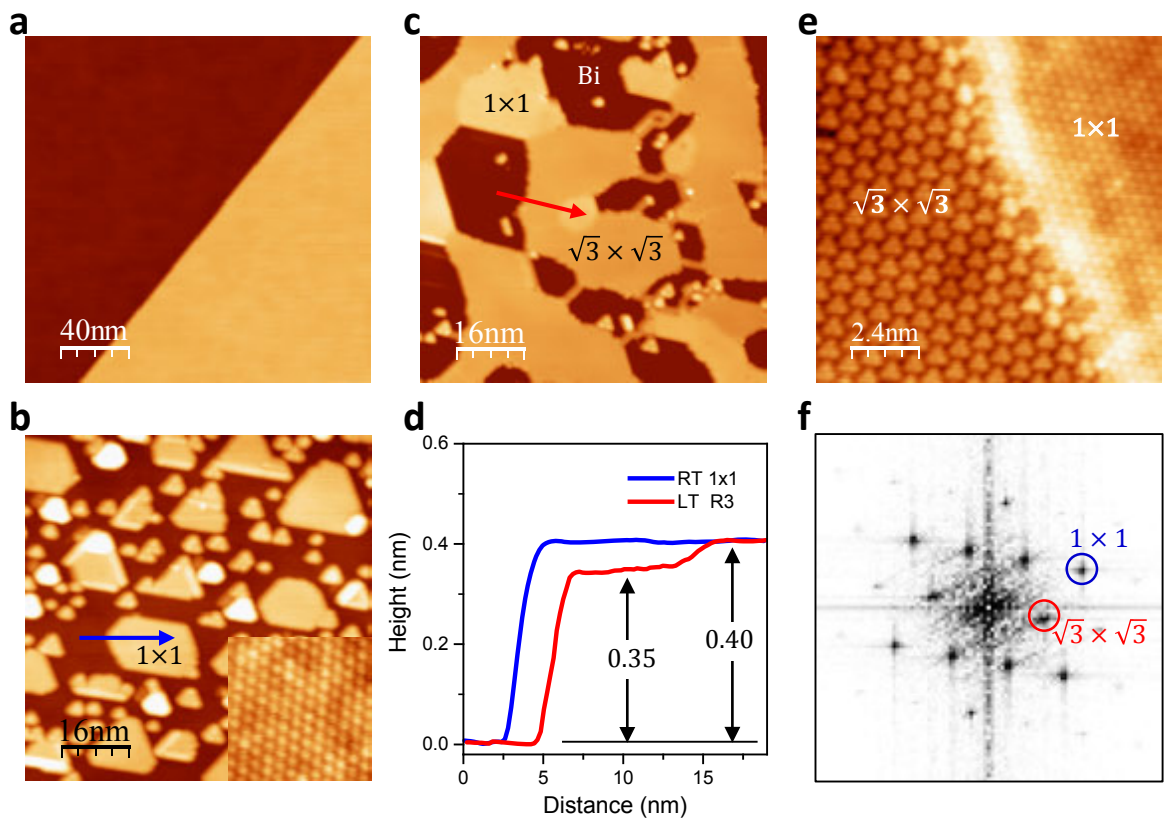

Figure 1

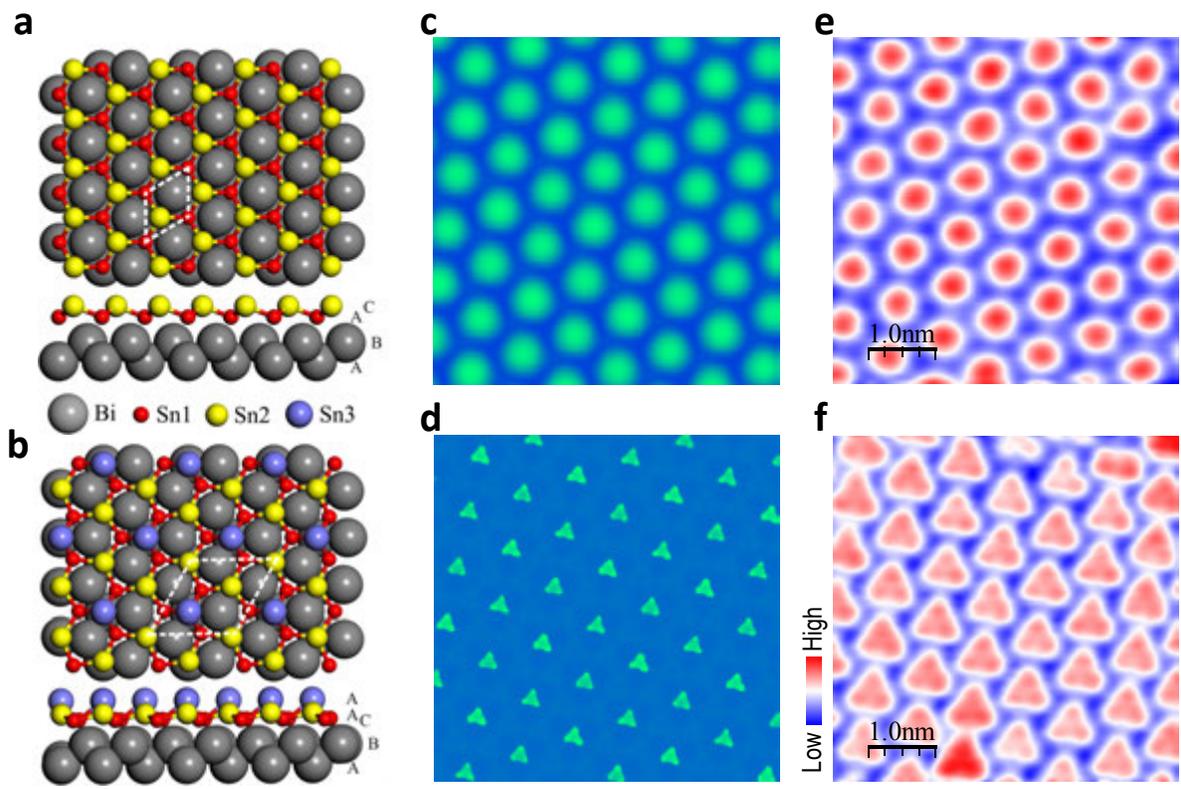

Figure 2

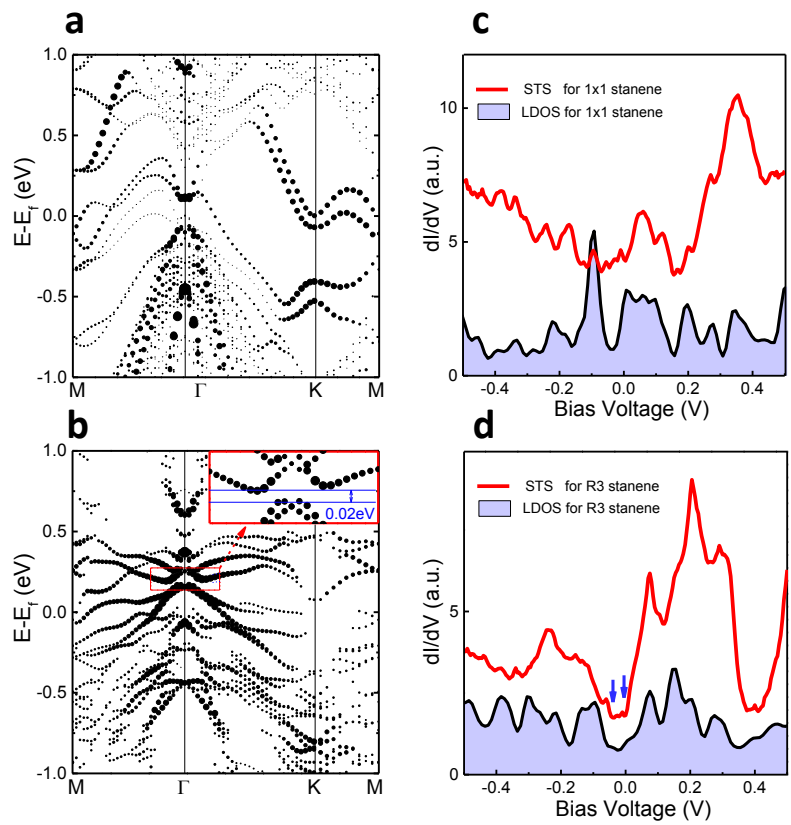

Figure 3

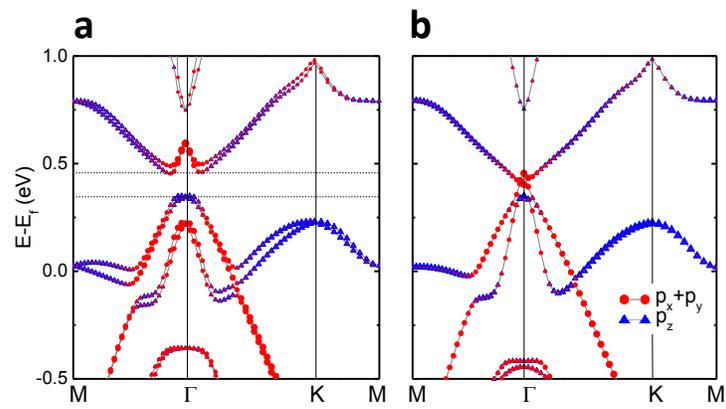

Figure 4

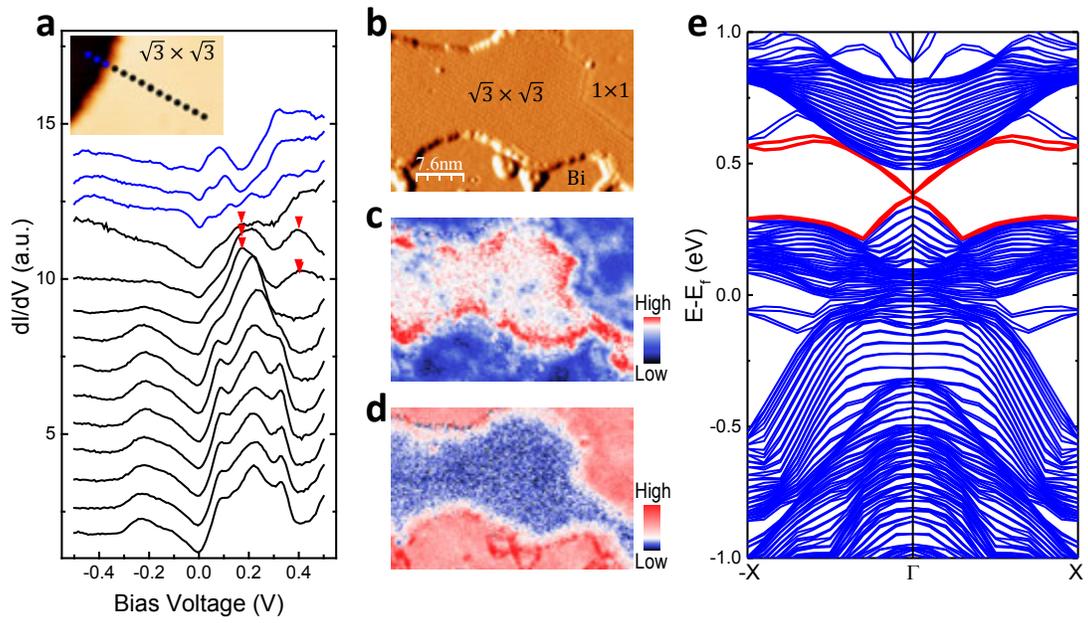

Figure 5